\documentclass{aa}
\usepackage{graphicx}
\usepackage{multirow}
\usepackage{txfonts}
\usepackage{floatflt}
\usepackage{longtable}
\usepackage[authoryear]{natbib}       
\bibpunct{(}{)}{;}{a}{}{,}
\newcommand{\kms}{km~s$^{-1}$}
\newcommand{\lam}{$\lambda$}
\newcommand{\KI}{\ion{K}{I}}

\newcommand{\CaI}{\ion{Ca}{I}}

\newcommand{\MnI}{\ion{Mn}{I}}
\newcommand{\VI}{\ion{V}{I}}
\newcommand{\TiI}{\ion{Ti}{I}}
\newcommand{\FeI}{\ion{Fe}{I}}
\newcommand{\FeII}{\ion{Fe}{II}}
\newcommand{\CrI}{\ion{Cr}{I}}
\newcommand{\BaII}{\ion{Ba}{II}}
\newcommand{\BaI}{\ion{Ba}{I}}
\newcommand{\LiI}{\ion{Li}{I}}
\begin{document}
\title{High-resolution optical spectroscopy of V838~Monocerotis in 2009
     \thanks{Based on observations made with ESO
    Telescopes at the Paranal Observatory under programme ID
    382.D-0152(C).}} 
\author{R. Tylenda\inst{1}  
\and T. Kami\'nski\inst{1,4}
\and M. Schmidt\inst{1} 
\and R. Kurtev\inst{2}
\and T. Tomov\inst{3}}
\offprints{R. Tylenda}
\institute{Department for Astrophysics, Nicolaus Copernicus
           Astronomical Center,
           Rabia\'{n}ska 8,
           87-100 Toru\'{n}, Poland\\ 
            \email{tylenda, tomkam, schmidt@ncac.torun.pl}
\and Departamento de Fisica y Astronomia, Universidad de Valparaiso,
     ave. Gran Bretana 1111, casilla 5030, Valparaiso, Chile\\ \email{radostin.kurtev@uv.cl}
\and Toru\'n Centre for Astronomy, Nicolaus Copernicus University,
     Gagarina 11, 87--100 Toru\'n, Poland\\ \email{Toma.Tomov@astri.uni.torun.pl}
\and currently at Max-Planck-Institut f\"ur Radioastronomie, Auf dem
            H\"ugel 69, 53121 Bonn, Germany
}

\date{Received; accepted}

\abstract
{V838 Mon erupted at the beginning of 2002. In the course of the outburst the
object evolved to low effective temperatures and declined as a very late
M-type supergiant. Among various scenarios
proposed to explain the nature of the outburst, the most promising is
a stellar merger event.}
{We aim at studying the structure and evolution of the object in the decline
from the 2002 eruption.}
{We obtained spectroscopic observations of V838~Mon in January--March~2009
with UVES/VLT. The results are analysed and compared with similar
observations obtained in October~2005 with HIRES/Keck.}
{The most striking difference between 2009 and 2005 is a complete absence of
the B3~V component and of the [\FeII] emission lines in 2009. The
present spectrum displays only the spectrum of the 2002 eruption remnant. 
It resembles that of
an $\sim$M6 giant, although the molecular bands in V838~Mon are deeper than
those in standard stellar spectra of a similar spectral class. Several
atomic lines, which displayed P-Cyg profiles in 2005, are now dominated by
pure absorptions. Some of these lines, however, show a narrow emission
component, which, as we argue, measures the radial velocity of V838~Mon. The
resulting heliocentric velocity is 71~\kms, which very well agrees with the
velocity obtained from SiO maser observations. The atomic lines and the
molecular bands show very complex kinematics. In some atomic lines and
high-excitation molecular bands we observe matter infalling in the object
atmosphere. The infall components were already observed in 2005, but were less
pronounced and present in fewer lines than in 2009. We argue that the most negative
radial velocity components seen in the resonance atomic lines and in the
low-excitation molecular bands were formed
in the ejecta of the 2002 eruption. The B3~V companion most probably became
engulfed in an opaque dusty matter of the 2002 V838 Mon ejecta.}
{}

        
\titlerunning{V838~Mon spectrum in 2009}
\authorrunning{Tylenda et al.}
\maketitle

\section{Introduction}
The eruption of V838 Mon was observed in January-March 2002
\citep[e.g.][]{muna02}. The object attracted significant public attention,
mainly because of its spectacular light echo \citep{bond03}. The eruption
itself was also very interesting and intriguing. The light curve
could have been taken for that of a slow nova, but the colour and
spectroscopic evolution of the object suggested a different nature. In the
course of the eruption the spectrum evolved to late spectral types and the
object declined as a very late M-type supergiant. At present, soon a decade after
the eruption, the remnant still shows a spectrum, which can be classified as
an $\sim$M6 giant or supergiant.

A few other stellar eruptions show characteristics very similar to those of
V838~Mon. They are named "V838~Mon-type objects" or "red novae".
Apart from V838~Mon, this class includes
M31~RV \citep{mould}, V4332~Sgr \citep{martini}, and V1309~Sco
\citep{mason10}. Some extragalactic intermediate-lumnosity optical
transients, like
M85\,OT2006 \citep{kulk07} and NGC300\,OT2008 \citep{bond09,berger09},
probably also belong to this class of stellar eruptions.

The nature of the V838 Mon-type eruptions has not yet been fully explained. 
As discussed in \citet{tylsok06}, thermonuclear mechanisms such as classical nova
\citep{tutu} or a late He-shell flash \citep{lawlor} should be
excluded. Instead, \citet{tylsok06} showed that all the main observational
characteristics of the
V838~Mon-type eruptions can be consistently understood as resulting
from stellar collisions and mergers. This idea obtained a strong support
from archive photometric observations of the progenitor of V1309~Sco
analysed in \citet{tyl_ogle}. The data evidently show that the progenitor
was a contact binary quickly evolving towards its merger.
According to \citet{kfs10} and \citet{ks10}, certain optical transients may result 
from mass transfer events in binary systems.  

A B3~V companion of V838~Mon was discovered when the main object declined 
from the 2002 eruption \citep{mdh02}. The companion remained constant in
brightness until an eclipse-like event in November--December 2006
\citep{muna07}. In 2004, narrow emission lines, predominantly from [\FeII], 
appeared in the spectrum of V838~Mon \citep{barsuk06}. The lines
strengthened with time and reached a maximum near the 2006  eclipse-like event
\citep{muna07}. The emission lines were most likely excited by the radiation
of the B3~V companion in the matter ejected by V838~Mon in 2002 that
approached the companion \citep{keckII}. The eclipse-like fading
of the companion in
2006 was accordingly caused by a cloud of dusty matter crossing the line
of sight of the companion \citep{keckII}.

A few months after the  eclipse-like event, the B3~V companion started a new
decline, this time slower but significantly deeper 
and more complex.\footnote{see http://jet.sao.ru/jet/$\sim$goray/.}
In contrast to the 2006 event, the second decline of the companion was also
accompanied by a decline of the [\FeII] emission lines. This indicated that the
radiation of the B3~V companion faded not only for the observer but also for
the matter flowing by the companion. Thus the event cannot be explained by
another dusty cloud crossing the line of sight. The companion was instead
engulfed in dense and opaque matter.
Two years after the beginning of the decline, last traces of the B3~V
companion disappeared from the spectrum of V838~Mon.

In March~2008, a field of V838~Mon was observed by XMM/Newton and a variable
X-ray source was detected near the position of the object \citep{anton}. The
X-rays may have resulted either from magnetic activity of the merger
remnant or from interactions of the 2002 ejecta with the B3~V companion. No
such source was detected in Chandra observations obtained a year after the
outburst, nor in January~2010.

In October 2005, we obtained a high-resolution spectrum of V838~Mon with 
the HIRES/Keck instrument \citep{keckI}. A detailed analysis of that data
was made in \citet{keckII}. In the present paper, we report on similar
observations that were obtained in January--March~2009 with the UVES/VLT
instrument.

\section{Observations}
\subsection{Spectroscopy}
V838 Mon was observed with the UVES spectrograph \citep{uves} on
the Very Large Telescope (VLT) in 2009 in three observing runs: on 
2~January, 13--14~February, and 22--23~March. For each run two exposures were
obtained with a total exposure time of 85 min. Technical details of
the individual exposures, i.e. observation date, UT time and Julian date at 
the exposure start, integration time, averaged airmass, 
and average seeing conditions calculated
over the exposure time, are given in Table~\ref{obslog}. The observations
were performed in the UVES standard {\it red arm 580} setting, which  gives
a spectral coverage of 4770--6825~\AA\ with an inter-chip gap between
5785~\AA\ and 5820~\AA. We used a slit with a width of 0.7 arcsec and the UVES 
image slicer no. 1. These settings provided a spectral resolution of 
about $\lambda/\Delta \lambda$=52\,000.   

All spectra were reduced  and calibrated in
IRAF\footnote{IRAF is distributed by the
  National Optical Astronomy Observatories,
which are operated by the Association of Universities for Research
in Astronomy, Inc., under cooperative agreement with the National
Science Foundation.} using standard procedures for Echelle
spectra \citep{irafech}. The reduction procedures included bias- and
flat-field corrections, removal of scattered light, and spectra extraction
with background subtraction. Cosmic ray events were rejected from the CCD
frames with the {\it Laplacian cosmic ray identification} algorithm
\citep{dokkum}. Observations were wavelength-calibrated using
spectra of Th-Ar lamps, and dispersion solutions were fitted to
identified lines with a typical accuracy of rms=3~m\AA. Flux
calibration was based on observations of the spectrophotometric
standard star Hiltner\,600 and its spectrophotometric fluxes tabulated
every 50~\AA\ in \citet{ham1,ham2}. We asses the overall accuracy of
the relative flux calibration to be about 5\%, but it is slightly poorer
within strong telluric absorption bands and in spectral regions
corresponding to the broad Balmer lines in the spectrum of the
standard star. The flux calibration
of the spectra from 24--25~March was based on standard-star
observations on 26~March and therefore the absolute flux calibration
of the March spectra is most uncertain. All spectra were shifted to
the heliocentric rest frame, and all Echelle orders from two
expositions obtained in the same run were merged into one spectrum. 
The signal-to-noise ratio reaches a value of 120 in the spectral region 
with the highest flux level.  

In order to check the accuracy of the absolute flux calibration of our
spectra, we estimated their corresponding $V$ magnitudes by
integrating the spectra convolved with the response curve of the
$V$ filter. The inter-chip gap was interpolated with the shape of the 
spectrum reported in \cite{keckI} as a reference. The magnitudes derived
from the spectra obtained in January, February, and March are
15.4, 15.5, and 15.3, respectively. They can be compared to the $V$ values of 
15.36 and 15.34 derived from
our photometric measurements made in December~2008 and March~2009, which we
described in Sect.~\ref{photo} (see Table~\ref{phottab}).

\begin{table*}\begin{minipage}[t]{\hsize} %
\caption{Technical details of the UVES/VLT observation of V838 Mon and the 
  spectrophotometric standard Hiltner 600.}\label{obslog}
\centering\renewcommand{\footnoterule}{} 
\begin{tabular}{ccccrcc}
\hline
date&UT&JD--24e5&object&\multicolumn{1}{c}{exp. time
  (s)}&airmass&seeing (\arcsec)\\
\hline
2009-01-02&  05:04:29&   54833.211&  V838 Mon&      2105   &   1.07 &	 1.53\\
2009-01-02&  05:56:33&   54833.248&  V838 Mon&      3005   &   1.10 &	 1.57\\	  
2009-01-02&  05:45:50&   54833.240&  Hilt 600&     305     &   1.17 &	 1.52\\
2009-02-13&  01:29:13&   54875.062&  V838 Mon&      3005   &   1.09 &	 1.57\\
2009-02-14&  03:54:36&   54876.163&  V838 Mon&      2105   &   1.19 &	 1.50\\
2009-02-16&  02:58:02&   54878.124&  Hilt 600&     305     &   1.18 &	 0.97\\
2009-03-22&  01:32:37&   54912.064&  V838 Mon&      3005   &   1.19 &	 0.86\\
2009-03-23&  01:25:30&   54913.059&  V838 Mon&      2105   &   1.18 &	 0.63\\
2009-03-23&  02:06:34&   54913.088&  Hilt 600&     305     &   1.48 &	 0.74\\
\hline\end{tabular}\end{minipage}\end{table*}

The three spectra from January, February, and March have essentially the same
shape. However, they are systematically shifted towards each other on the flux
scale. The flux level in the spectrum from February is on average
7$\pm$2\% lower than that of the spectrum from January, and the last spectrum from
March is 14$\pm$2\% brighter than the spectrum from
January. These flux variations are within the accuracy of the
absolute flux calibration of our spectra. 
We can conclude that the object had essentially the same flux level in
the three observations.
Minor differences between the three spectra related to individual features
are discussed in Sect.~\ref{var}.
We have summed up the three spectra and  presente
and analyse the resulting spectrum.

\subsection{Photometry  \label{photo}}
V838 Mon was measured in the optical photometric bands several times in 2008--2009. 
On 18~December~2008 the object was observed in the $V$ and $I_C$ filters using a CCD camera at 
the 1~m~telescope of the South African Astronomical Observatory (SAAO). 
Photometric observations
were also carried out on 16~March~2009 in the $BVR_CI_C$ filters with 
the 1~m~SMARTS telescope of the Cerro Tololo Inter-American Observatory (CTIO). 
Finally, a $BVI_C$ photometry was performed on 5--6~October~2009 
with the 1~m SAAO telescope. 
  
All the above observations were reduced in IRAF using standard procedures. 
The magnitudes of V838~Mon were derived using photometry of field stars from 
\citet{muna02}. It should be noted, however, that the quality of the photometry 
in Munari et al. was questioned in \cite{bond07}, which could introduce some extra 
systematic errors to our photometry. 
Magnitudes corresponding to different exposures from the same observing night 
were averaged for each filter. The results are given in Table~\ref{phottab}, 
where errors are also provided; they include the 3$\times$rms (root mean square) 
scatter in magnitudes of different exposures during the night and the 3$\times$rms 
uncertainty in the transformation factors. In 2008 V838~Mon became the brightest 
$I_C$ source in the field of Munari et~al., making its calibrations 
uncertain in this band. 

Near-infrared (NIR) photometric observations of V838~Mon in the $JHK_S$ bands 
were carried out on 13~February~2009 with the CAIN-3 camera installed on 
the 1.52~m~Carlos S\'{a}nchez Telescope (TCS)\footnote{The 1.52m Carlos
S\'{a}nchez Telescope is operated on the island of Tenerife by the Instituto de
    Astrofisica de Canarias in the Spanish Observatorio del Teide.}.
The date of the NIR observations 
is the same as that of the second spectroscopic observing run at UVES/VLT. 
Data were reduced using CAIN data reduction 
scripts\footnote{http://www.iac.es/telescopes/cain/reduc/caindr.html} developed 
in IRAF by the Instituto de Astrof\'{i}sica de Canarias. 
The instrumental magnitudes were converted to the standard system using observations 
of two standard stars FS\,11 and FS\,18. Their NIR magnitudes were taken from a list of the UKIRT faint 
standard stars \citep{ukirt}. The resulting averaged NIR magnitudes for V838 Mon are 
presented in Table~\ref{phottab}. 

\begin{table}\begin{minipage}[t]{\hsize} 
\caption{Results of the photometric measurements of V838 Mon in 2008--2009.}\label{phottab}
\centering\renewcommand{\footnoterule}{} 
\begin{tabular}{cccccc}\hline
Date&JD&Band&Mag&Error&Observat.\\
\hline
18 Dec. 2008&2454819.45&$V$  &15.36&0.02&SAAO\\
18 Dec. 2008&2454819.45&$I_C$&10.08&0.10&SAAO\\
13 Feb. 2009&2454876.40&$J$  &7.43&0.13&TCS\\
13 Feb. 2009&2454876.40&$H$  &6.25&0.24&TCS\\
13 Feb. 2009&2454876.40&$K_S$&5.50&0.07&TCS\\
16 Mar. 2009&2454906.52&$B$  &19.11&0.13&CTIO\\
16 Mar. 2009&2454906.53&$V$  &15.34&0.05&CTIO\\
16 Mar. 2009&2454906.54&$R_C$&12.63&0.06&CTIO\\
16 Mar. 2009&2454906.55&$I_C$&10.28&0.12&CTIO\\
05 Oct. 2009&2455109.64&$B$  &18.58&0.66&SAAO\\
05 Oct. 2009&2455109.63&$V$  &15.74&0.28&SAAO\\
05 Oct. 2009&2455109.63&$I_C$&10.02&0.15&SAAO\\
06 Oct. 2009&2455110.63&$B$  &18.69&0.43&SAAO\\
06 Oct. 2009&2455110.62&$V$  &15.65&0.39&SAAO\\
\hline\end{tabular}\end{minipage}\end{table}

\section{Results, analysis, and comparison to the spectrum obtained in October 2005}

\subsection{Spectral features  \label{spect_features}}

\begin{figure*}
\includegraphics[angle=270, width=\textwidth]{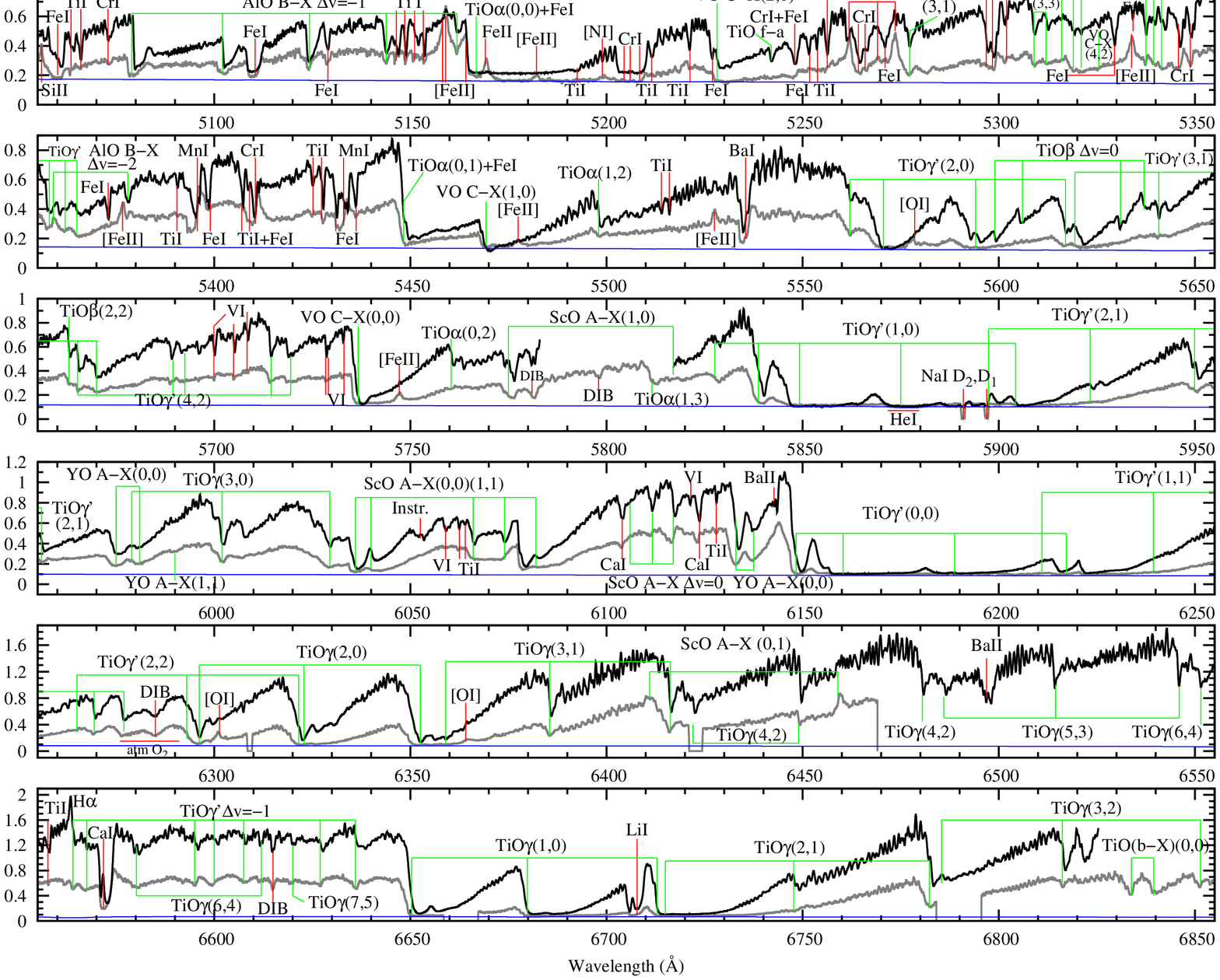}
\caption{Average spectrum obtained with UVES/VLT in January--March~2009 (black line) 
is compared to the spectrum obtained with HIRES/Keck (grey line) in October~2005. 
The spectra are deredenned with $E_{B-V}=0.9$ and $R_V=3.1$, and smoothed. 
A synthetic spectrum of a B3~V star, which contributes to the 2005 fluxes, is shown 
for comparison (blue line). The UVES spectrum was shifted in fluxes by the values 
corresponding to the average contribution of the B3~V companion in the 2005 spectrum; 
the value of the shift is 0.21 for the two top panels, and 0.10 for the other 
panels. Identified atomic spectral features are indicated by red markers, 
while molecular band-heads are assigned green markers. 
The ordinate units are 10$^{-13}$~erg~s$^{-1}$~cm$^{-2}$\AA$^{-1}$.}  
\label{atlas}
\end{figure*}

A high-resolution spectrum of V838 Mon in its
post-outburst phase was presented in \citet{keckI} and analysed in
\citet{keckII}. This was a HIRES/Keck spectrum obtained in October~2005, i.e. a
year before the eclipse-like event observed in November/December~2006
\citep[see, e.g.][]{muna07}. A detailed atlas of
spectral features in that spectrum can be found in \citet{keckI}. Most of
the identified features there are also present in the 2009 UVES/VLT
spectrum analysed in this paper. The 2009 spectrum is compared to the
2005 one in Fig~\ref{atlas}.

The complete absence of the spectrum of the B3~V companion and of the
[\ion{Fe}{II}] emission lines in the 2009 spectrum is the most striking 
difference between the two spectra. These spectral components
were best observed in the blue part of the 2005 spectrum, which, i.e.
$\lambda < 4750$\,\AA, was not observed in 2009. Nevertheless, from the
available data we can quite firmly conclude that the B3~V component disappeared.
Firstly, in 2009, we do not see any trace of the broad H$\beta$ absorption line, which
was prominent in 2005 (see Fig~\ref{atlas}). Secondly, in the deep, heavily
saturated TiO bands, i.e. at $\sim$4950\,\AA, 5850--5910\,\AA\ or 6160--6230\,\AA, 
we saw a residual flux in 2005, which was nicely intepreted as the continuum from the
B3~V companion. In the 2009 spectrum the flux in these regions goes practically to a zero
level, i.e. significantly below the 2005 level (see Fig.~\ref{atlas}).
Comparing the observed fluxes in these regions, we can conclude that in 2009 the B3~V
spectral component was at least 30 times fainter than in 2005.

As mentioned above, the emission features, which were present in 2005, 
are not seen in the 2009 spectrum.
We have not found any remnant emissions at the positions of the [\ion{Fe}{II}]
lines, which were strong in 2005. Comparing the observed fluxes in the
regions of the strongest [\ion{Fe}{II}] emission lines in 2005, we can
conclude that these lines in 2009 are at least 20 times fainter than in
2005.

The emission components of the
resonance lines showing P-Cyg profiles in 2005 also 
became very faint or disappeared completely
in 2009. Similarly, the emission band of
ScO near 6460~\AA\ is not seen now. The only emission feature, which is
clearly present in the 2009 spectrum, is that of
the H$\alpha$ line (see Figs.~\ref{atlas} and \ref{profiles}d).

Almost all the atomic absorptions found in 2005 are
present in the 2009 spectrum. Their profiles, however, changed
considerably (see Sect.~\ref{sect_profiles}). In addition, the present spectrum 
displays several absorption features of
\FeI, \TiI, \VI, and \CrI, which were very weak or absent in 2005. These
lines belong to multiplets with lower energy levels of
$\la$1.1~eV.

In 2009, V838 Mon shows a very rich molecular spectrum in absorption.
Practically, all the features identified in the atlas of \citet{keckI} are
present in the UVES spectrum. Some of the bands are more clearly seen in the
2009 spectrum, which allowed us to improve their identification. The shapes and strengths 
of some of the bands have changed and this is described in detail in
Sect.~\ref{bands}. In general, many of the strongly saturated bands of TiO are 
now seen to be less saturated.

Certain spectral regions of the UVES spectrum were not covered
in the 2005 spectrum. Identifications of spectral features in 
these regions can be found in Fig.~\ref{atlas}. These ranges give access 
to two strong lines of \ion{Ba}{II}\,\lam\lam\,4934 and 6496, and absorption 
bands of AlO (the B--X system) and TiO (the $\gamma'$ system).      

\subsection{Spectral energy distribution}\label{sed}

We compared the 2009 spectrum with UVES spectra of red giants 
from the Paranal Atlas \citep{paranal}. When the whole spectrum is taken into account, 
the best fitting spectra have types of M6$\pm$2. 
Although the reference spectra 
reproduce the spectrum of V838 Mon quite well in the regions with a strong 
pseudo-continuum, numerous absorption bands are significantly deeper in 
V838~Mon. The same was observed in 2005 \citep{keckII}.

As can be seen from Fig.~\ref{atlas}, V838~Mon in 2009
was significantly brighter than in 2005.
After removing the flux 
of the B3~V star from the 2005 spectrum, we found that the flux of
V838~Mon itself in the observed spectral region 
increased by a factor of $\sim$2 between 2005 and 2009. This growth in
the optical flux was caused by slight increase of V838~Mon in the effective
temperature and luminosity (see below), as well as a lower blanketing in
the molecular bands in 2009 (see Sect.~\ref{bands}).

\begin{figure}
  \resizebox{\hsize}{!}{\includegraphics{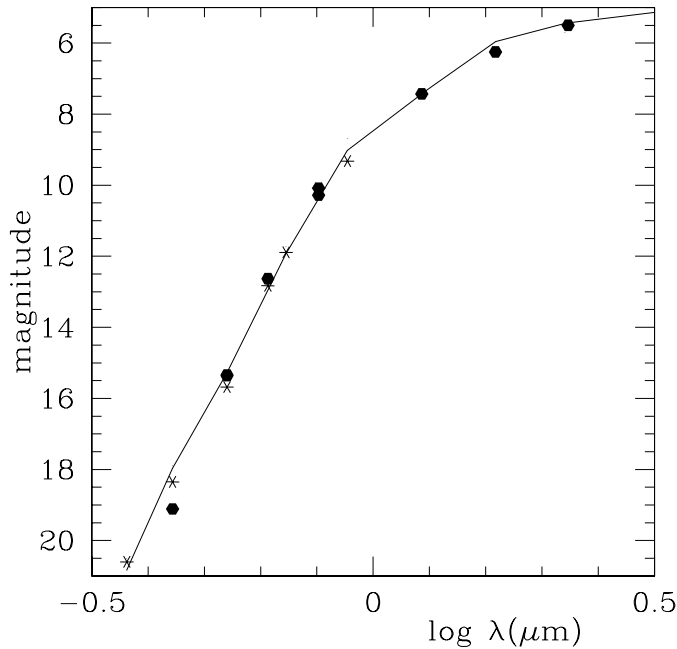}}
\caption{Standard supergiant photometric spectrum (M6.3 reddened with
$E_{B-V}=0.9$: full line) fitted to the photometric results obtained
in December~2008 -- March~2009. Full points:
our data presented in Table~\ref{phottab}, asterisks: mean values from
the data compiled by V. Goranskij.
See text for more details.}
\label{sed_fig}
\end{figure}

We fitted standard broad-band photometric spectra to our photometric
measurements obtained in December~2008 -- March~2009 
 supplemented with mean values taken from the
measurements obtained in the same time period that were compiled by
V.~Goranskij$^1$
\citep[for details of the fitting
procedure see][]{tyl05}. Figure~\ref{sed_fig} shows
the best fit of supergiant (luminosity class I) standards reddened 
with $E_{B-V}=0.9$ to the photometric data. The
spectral type of the fitted supergiant is M6.3, which agrees well with the
spectral type  from
comparing the Paranal spectra to that of V838~Mon, derived above.
This spectral type (M6.3\,I) corresponds to an
effective temperature of $\sim$3270~K \citep{levesque}. For a distance of
6.5~kpc \citep{bond07,kam07,sparks08} the object has an effective radius of 
$\sim$380~$R_\odot$ and a
luminosity of $\sim$1.5\ 10$^4~L_\odot$.
Practically the same parameters are obtained when fitting standard giant 
(luminosity class III) photometric spectra. 
When compared to the results of a similar
analysis made in \cite{keckI}\footnote{Note that in \citet{keckI} the
effective temperature was derived following the spectral type vs. effective
temperature calibration of \citet{sk82}.
With the calibration of \citet{levesque} the effective temprature in 2005
would have been $\sim$3200~K.}, in 2009 V838~Mon was slightly hotter and
$\sim$20\% more luminous than in 2005.

\subsection{Profiles of atomic lines}\label{sect_profiles}
\begin{figure*} \centering
\includegraphics[width=14cm]{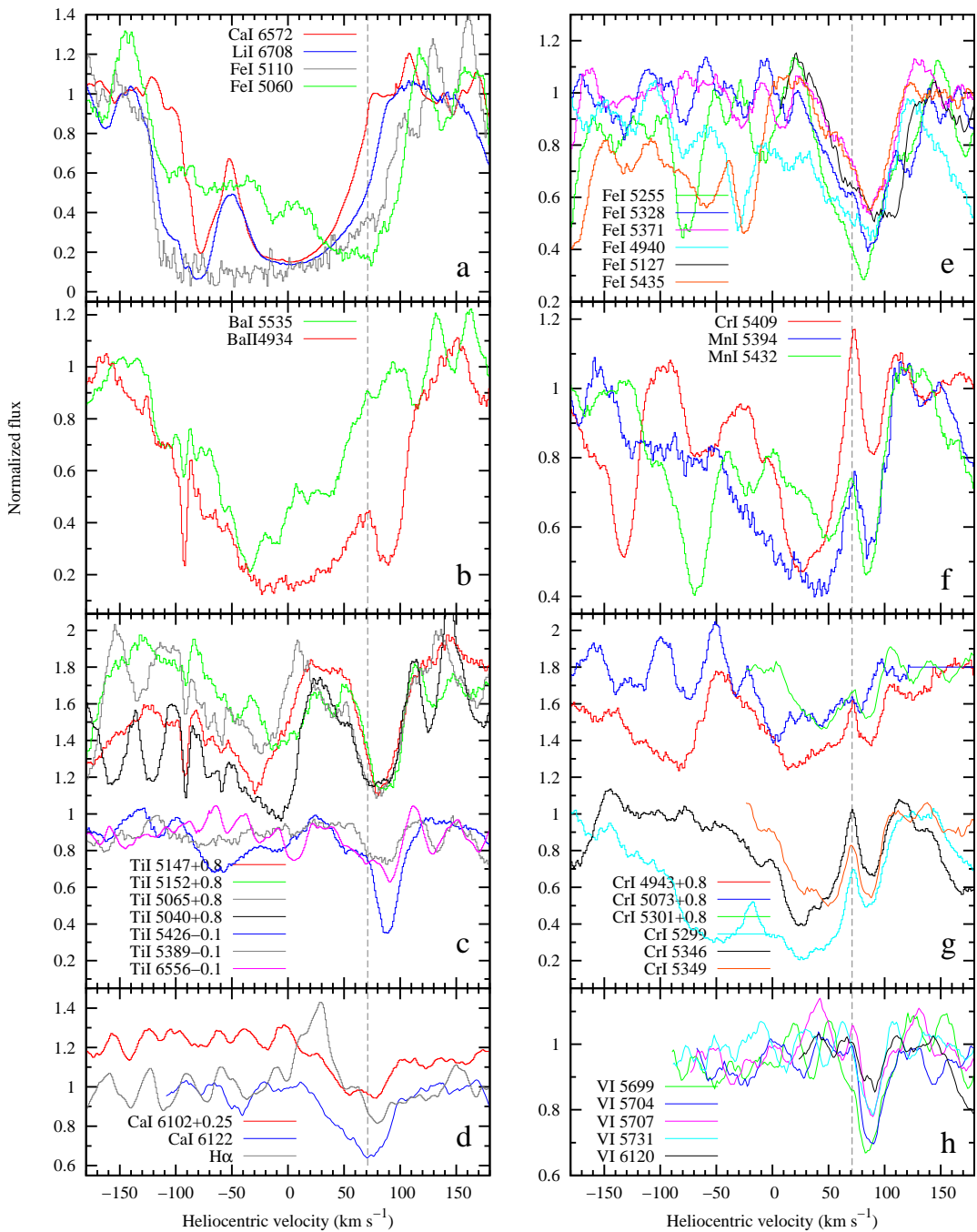}
\caption{Sample of profiles of atomic lines in the average UVES spectrum from 2009. 
The vertical line marks the velocity of the maser of V838 Mon. Some profiles were 
shifted in ordinate for more clarity. Many structures, particularly seen at
negative radial velocities, are caused by blending molecular bands.}  
\label{profiles}
\end{figure*}

The atomic lines observed in our spectrum show complex structures in their 
profiles. Examples are shown in Fig.~\ref{profiles}. 
In the following, we assume that the stellar radial 
velocity is equal to the value derived form SiO masers
\citep{deguchi,claussen}, i.e. $V_{\rm h}\simeq71$~\kms\, shown with a dashed line
in Fig.~\ref{profiles}. This point is further discussed in
Sect.~\ref{em_comp}. All radial velocities given
in this paper are in the heliocentric rest frame.

As discussed in \cite{keckI}, many resonance atomic lines in the spectrum of 
V838~Mon taken in October~2005 showed P-Cyg profiles. All these lines
(except \KI\,\lam\,7698, \ion{Rb}{I}\,\lam\,7800, and
\ion{Mg}{I}\,\lam\,4571, which are outside the wavelength range of the
present spectrum)
are observed in the 2009 spectrum, but no more as P-Cyg ones; their emission
components practically disappeared. The absorption component in
many lines is now wider. It is usually more extended towards longer
wavelengths, which is at least partly due to the disappearance of the emission
component. Some absorptions are shifted to shorter wavelengths,
but never beyond the maximum blueshifted absorption observed in 2005. 
The maximum outflow velocity of $\simeq 215$~\kms,
derived from the 2005 spectrum \citep{keckII}, remains practically the same in
2009.

The absorption lines show complex profiles in most cases. 
A relatively narrow absorption centred at $V_{\rm h}\simeq87$~\kms\ 
(see Fig.~\ref{profiles}) is one of the most
characteristic features seen in the majority of the lines.
In the 2005 spectrum, this component was seen in the \TiI\ lines
\citep{keckI} and was interpreted as a signature of an infall in the
V838~Mon atmosphere \citep{keckII}. The wings of this feature in the
\TiI\ lines are now observed at practically the same velocities as in
2005, but the feature is now deeper and the maximum absorption is at a slightly
higher velocity.

The infall component is the only feature seen in the \VI\ lines (see
panel h in Fig.~\ref{profiles}). These are lines arising from levels with
excitation energies as high as $\sim$1.05\,eV. When fitted with a Gaussian,
the component has 
a full-width-at-half depth (FWHD) of $\sim$15~\kms. We note that the \VI\
lines were also seen in the 2005 spectrum, but they were then significantly
weaker and their profiles were more noisy. Therefore, these lines were not analysed 
in \cite{keckI} and \cite{keckII}.
Comparing the 2005 spectrum with the present one we can now state that in
2005 the \VI\ lines were of similar width but at slightly lower radial
velocities than in 2009. This is consistent with the analysis
of the infall component in the \TiI\ lines, presented above.

The profiles of non-resonance \FeI\ lines (panel e in Fig.~\ref{profiles})
show a maximum absorption at velocities consistent with the infall. However,
they are wider than the profiles of the \VI\ lines and considerably
asymmetric. The blue wing extends well below the stellar velocity,
indicating an outflow. Contrary to resonance \FeI\ lines (displayed in
panel a in Fig.~\ref{profiles}) the non-resonance lines do not show any
significant feature at $V_{\rm h} \la 20$~\kms\ (structures seen in panel e
of Fig.~\ref{profiles} at these velocities are caused by molecular bands).

The infall component in the lines of \MnI\ and \CrI\ (panels f and g in
Fig.~\ref{profiles}) is separated from the blueshifted (indicating
outflow) part of the absorption profile by a narrow emission-like feature peaking at
the stellar velocity. In Sect.~\ref{em_comp} we show that this is most likely a real
emission feature.

Clear outflow components, i.e. absorptions at $V_{\rm h} \la 50$~\kms\, are 
primarily seen in
lines arising from levels of low excitation energy, i.e. ground states 
(\CaI, \LiI, \FeI: panel a in Fig.~\ref{profiles}; \BaI,
\BaII: panel b; \MnI: panel f) and levels with energies $\la$0.05\,eV (\TiI\
lines in panel c)\footnote{There is a mistake in \cite{keckI} and
\cite{keckII}, where the upper energy level of the \TiI\ lines was given 
as the lower one.}.
They are also seen in the  \CrI\ lines (panel f and g), however, 
which arise from levels of energies as high as $\sim$1.0\,eV, i.e.
excitation energies similar to those of the \VI\ lines, which do not show
any outflow (panel h).

In the 2005 spectrum, a relatively narrow absorption component (NAC) was observed
at $V_{\rm h} \simeq -82$~\kms\ \citep[see Fig.~6 in][]{keckI}. 
It was particularly pronounced in
\ion{Rb}{I}\,\lam\,7800 and \ion{Mg}{I}\,\lam\,4571, but also present in
\MnI\,\lam\,5394 and 5432. The two former lines are, unfortunately, outside
the spectral range of the present spectrum, but the \MnI\ lines do not show
any clear feature at the above velocity in 2009. However, an absorption 
component at practically the same velocity is now observed in \CaI\,\lam\,6572 
and \LiI\,\lam\,6708 (see panel a in Fig.~\ref{profiles} and
Fig.~\ref{monthly}). In 2005, these two lines presented rather smooth
absorption profiles extending down to $V_{\rm h} \simeq -120$~\kms.
The 2009 component is 
 wider, however, than the similar feature in
2005, particularly in \LiI\,\lam\,6708. 

In the 2009 spectrum, we can easily identify a very narrow absorption
component at $V_{\rm h} \simeq -92$~\kms. It is well seen in the \BaI, 
\BaII, and \TiI\ lines (panels b and c in Fig.~\ref{profiles}), but a trace
of it seems to be also present in the blue wing of the wider feature in 
the \CaI\ and \LiI\ lines (panel a in Fig.~\ref{profiles}).
This very narrow feature has a Gaussian profile with 
FWHD\,=\,5.6~\kms, which is the same as the FWHM of the instrumental profile.
A similar feature was observed in 2005 in the \TiI\ lines, but at 
$V_{\rm h} \simeq -90$~\kms. It was also wider than now, but mostly because
of the lower resolution of the 2005 spectrum.

The \CaI\,\lam\,6102 and 6122 lines arise from the highest energy levels 
($\sim$1.9~eV)\footnote{There is a mistake in \cite{keckI}, where the upper energy level 
of these \CaI\ lines was given as the lower one.} among 
the lines identified in the 2009 spectrum. Consequently, they are likely to be formed 
deeply in the photosphere. This is supported by their central positions being close 
to the stellar velocity (see panel d in Fig.~\ref{profiles}). Their wings,
extending from $V_{\rm h} \simeq 10$~\kms\ to $V_{\rm h} \simeq 100$~\kms\ ,
show an outflow and infall in the atmosphere of V838~Mon.

The only clear emission line seen in the 2009 spectrum is that of H$\alpha$
(see panel d in Fig.~\ref{profiles}).  
The observed position of the line gives $V_{\rm h} \simeq 29$~\kms. 
The line is narrow with FWHM\,$\simeq 25$~\kms. 
The approximate line equivalent width (EW) and dereddened flux are 
$-0.24$~\AA, and $3.3 \times 10^{-14}$~erg~cm$^{-2}$s$^{-1}$, respectively. 
This feature was absent from the 2005 spectrum.

\subsection{Variability on a time-scale of months}\label{var}
\begin{figure}
\includegraphics[angle=270, width=\columnwidth]{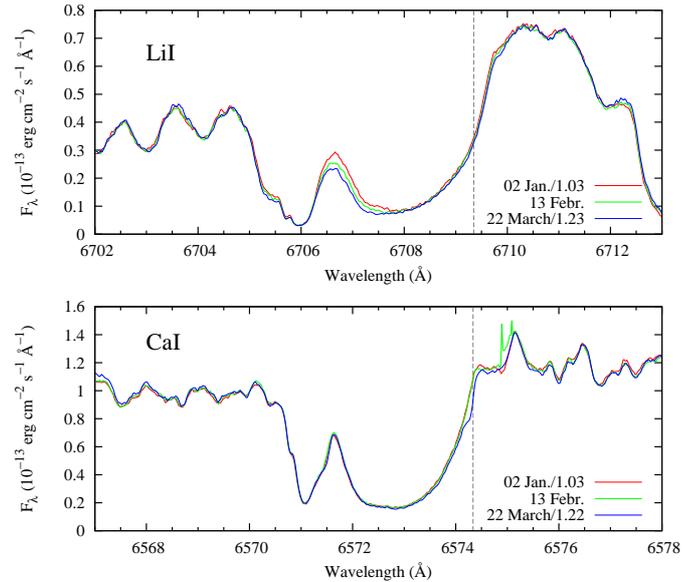}
\caption{Spectral variability in the profiles of the \LiI~\lam~6707 and 
\CaI~\lam~6572 lines as seen in our three UVES spectra from 2009. 
The January and March spectrum were rescaled to fit the local pseudo-continuum 
in February. The feature at around 6575~\AA\ in the February spectrum
caused by a cosmic-ray hit. The vertical dashed line marks the object
heliocentric velocity of 71~\kms.} 
\label{monthly} 
\end{figure}

A characteristic flow time-scale of the outflow in V838 Mon is 
$\tau \simeq R_{\star}/V_{\rm exp} \simeq 20$~days 
(for $R_{\star} \simeq 400$~R$_{\sun}$ 
and $V_{\rm exp} \simeq 150$~\kms), which is approximately the time span between 
our three observing runs. Certain differences 
in individual line profiles can be noticed between these three spectra. 
The changes are most pronounced in the \LiI~\lam~6707 line, which is expected 
to have the lowest optical depth among the strong absorption lines. 
The variability in \LiI\ line is shown in the top panel of Fig.~\ref{monthly}. 
The feature between the deep absorptions appears to slightly
decrease with time. Interestingly, the profile of 
the intercombination line of \CaI~\lam~6572, which exhibits
a similar shape as \LiI~\lam~6707, has not undergone similar changes; 
it shows instead only a minor change in the March spectrum when 
the absorption profile at around the maser velocity became slightly deeper 
than in the two earlier epochs (see bottom panel in Fig.~\ref{monthly}).  
Other lines do not show any clear signs of variability above their local noise 
levels.

\subsection{Emission in atomic lines at the stellar velocity \label{em_comp}}
The profiles of many lines displayed in Fig.~\ref{profiles} show an
emission-like component at the radial velocity of the SiO maser (71~\kms). This is
particularly the case for the \MnI\ and \CrI\ lines. This can be a real
emission, in some sense a remnant of the P-Cyg emission component observed in
2005. Another possibility is that the whole profile is purely absorptive,
but the higher flux near the maser velocity simply means that there is
little absorbing matter at these velocities. 

The case of the
\CrI\,\lam\,5409 (see panel f in Fig.~\ref{profiles}), where the
emission-like component peaks slightly above the continuum level, argues in
favour of the former possibility. Additional arguments arise from
comparing profiles of pairs of lines with the same lower energy
level, but different oscillator strengths (transition probabilities).
This is the case, for instance, for \CrI\,\lam\,5346 and \lam\,5349. Both
lines have the same lower level, but the former one has a $gf$-value
twice higher than the latter. In the case of pure
absorption, the \lam\,5346 line is expected to be deeper than the \lam\,5349 one.
Panel g in Fig.~\ref{profiles} shows that this is indeed the case except for
radial velocities $\la$50~\kms. For higher velocities, in particular in the vicinity
of the maser velocity, the situation is reversed, i.e. the flux in the \lam\,5346
line is higher than in the the \lam\,5349 one. The emission-like component
is clearly stronger in the former line. This shows that the
emission-like component at $V_{\rm h}$~=~71~\kms\ is indeed of emission origin.
The same conclusion can be drawn when comparing the \MnI\,\lam\,5394 and
\lam\,5432 lines (panel f in Fig.~\ref{profiles}). Both lines have the
ground level as their lower level, but the former one has a $gf$-value almost
three times higher than the latter.
Even the profiles of \FeI\,\lam\,5110 and \lam\,5060 (see panel a in
Fig.~\ref{profiles}),
although they do not display any clear emission component, show that the
high-velocity regions are affected by emission. These two lines arise from
the ground state, but the \lam\,5110 has a transition probability almost 40
times higher than the \lam\,5060. For $V_{\rm h} \la 50$~\kms\ the former line is
considerably deeper than the latter one, but for $V_{\rm h} \ga 50$~\kms\ the
opposite is observed.

We can state, therefore, that an emission
component is observed in some atomic lines. This is probably a residual emission 
component of the P-Cyg
profiles observed in 2005 and is most likely produced
in the moving (outflowing and/or infalling) matter above the photosphere
that is
observed above the photospheric limb. The extent of the emitting region is
probably small compared to the photospheric radius and therefore, we mainly
observe matter moving perpendicularly to the line of sight in the
emission component. Hence, the
emission is expected to be observed at radial velocities close to the radial
velocity of the object.
Indeed, the observed emission component
is relatively narrow and peaks at a velocity very close to that of the SiO
maser (Fig.~\ref{profiles}). 

This is the first reliable and precise measurement of the radial velocity of
V838~Mon obtained from optical spectroscopy.
The result agrees very well with that derived from the
SiO maser observations. In conclusion, the heliocentric radial velocity of
V838~Mon is $V_{\rm h}$~=~71~\kms. 


\subsection{Molecular bands  \label{bands}}

The molecular absorption features in the presented spectrum
are very similar to those observed in 2005 \citep{keckI}. We identified
bands of TiO, VO, ScO, YO, and AlO. Most of the
molecular bands are significantly deeper than in standard stars of a similar
spectral type to V838~Mon. However, in the present spectrum, they are mostly 
observed to be weaker than in the 2005 one. 
Similarly as in \cite{keckI}, when interpreting the observed band
shapes, simulations of a homogeneous absorbing
layer situated above the stellar photosphere were performed.
Usually, a velocity dispersion of 10--20~\kms\ has to be adopted in the
simulations to reproduce the shapes of individual absorption
features.
Therefore, estimates of the radial velocity from the bands have a
typical uncertainity of $\sim$10~\kms. We do not present detailed 
figures comparing the results of the simulations with the observed profiles, 
since they are similar to those shown in \cite{keckI}. 
Detailes of the molecular bands, for which we were able to derive radial
velocities, are given in Appendix~\ref{table_molec}.

\subsubsection{TiO  \label{tio_sect}}

\begin{figure}
  \resizebox{\hsize}{!}{\includegraphics{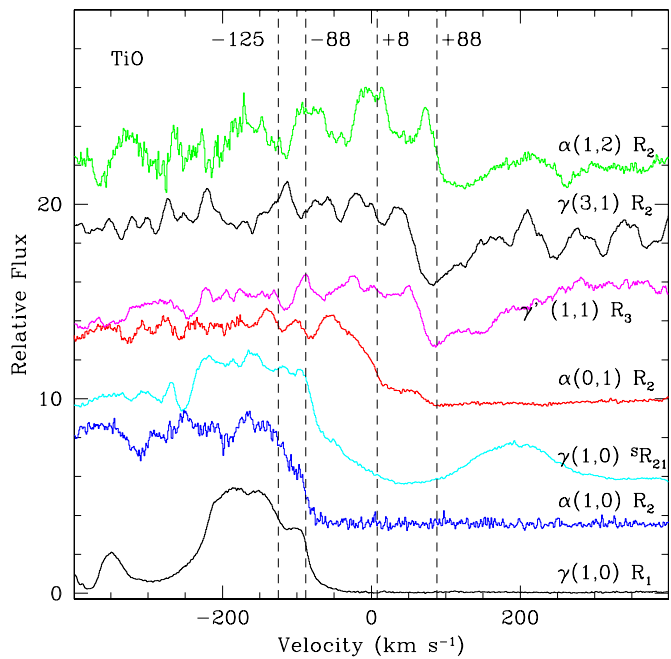}}
  \caption{
Sample of the TiO band-heads illustrating  the observed velocity components
marked with dashed vertical lines. The radial velocity is in the
heliocentric frame.
}
  \label{tio_fig}
\end{figure}

As can be seen from Fig. \ref{tio_fig}, different TiO bands show different
radial velocities. From the bands with relatively sharp heads
we were able to identify four velocity components,
i.e. --125, --88, 8, and 88~\kms, which are marked in Fig.~\ref{tio_fig}.
A general trend is very similar to that observed in \citet[][see their
Figs.~3 and 4]{keckII}, namely
that bands from low excitation levels are seen at low (negative)
velocities, while the absorptions from excited vibrational levels show high
(positive) velocities. The only important difference is that the highest
radial velocity is now observed to be $\sim$88~\kms, instead of
$\sim$58~\kms\ as in 2005.

\begin{figure}
  \resizebox{\hsize}{!}{\includegraphics{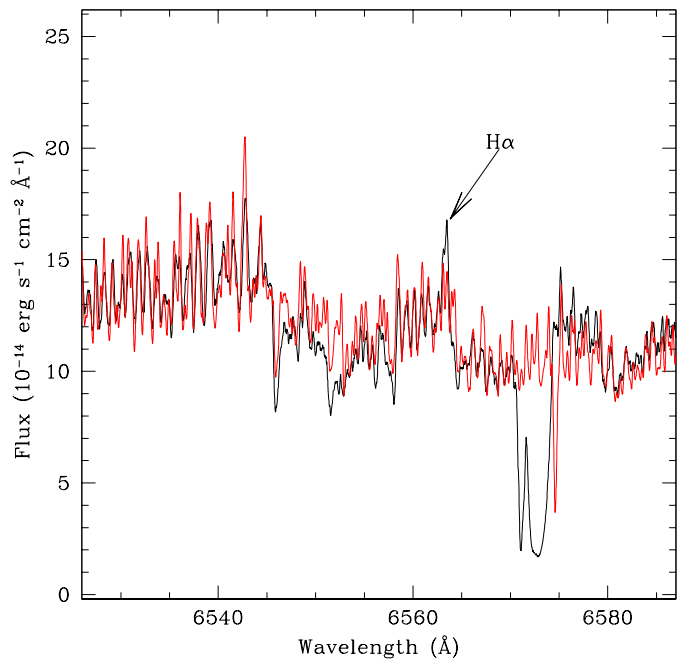}}
  \caption{
Part of the 2009 spectrum showing details of highly excited
rotational lines  of the TiO $\gamma$ (5,3) and (6,4) bands. 
The V838~Mon spectrum (in black) is overlaid with an M5~III spectrum (in red).
The latter spectrum was shifted to a heliocentric radial velocity of 88~\kms.
A narrow H$\alpha$ emission line and a broad \CaI\ absorption profile can be
recognized in the spectrum of V838 Mon. 
}
  \label{tio_lines_fig}
\end{figure}

The most negative component, i.e. --125~\kms, is the same as in 2005
\citep{keckI} and is observed
only in bands absorbing from the ground vibrational state of the
TiO system.
The --88~\kms\ component is seen in the bands absorbing from the ground
vibrational level of TiO, as well as of VO C-X (0,0) and ScO A-X (0,0).
The component at 8~\kms\ is particularly well seen in the AlO bands and is discussed below.

The most redshifted component at 88~\kms\ is observed in the excited bands, not only in
their heads, but also in their resolved rotational lines. An example is shown
in Fig.~\ref{tio_lines_fig}, where a portion of the V838~Mon spectrum is
overlaid with a standard M5~III spectrum \citep[taken from][]{paranal}.
An excellent fit of the positions of the TiO rotational lines
in both spectra was obtained
when the comparison spectrum was shifted to a heliocentric radial velocity of 
88~\kms. 
This velocity is observed in practically all highly-excited
(from the higher vibrational levels or high-J levels)
and weak absorption bands of TiO. Note that this is the same velocity as
that of the infalling components observed in numerous atomic lines,
as discussed in Sect.~\ref{sect_profiles}.

Similarly as in \cite{keckI}, an estimate of the excitation temperature of molecular
bands can be obtained from fitting the results of our simulations 
to the observed band shapes.
The shape of the TiO $\gamma$ (1,0) band
can be well reproduced by a layer of gas excited
to $\sim$500 K. To reconstruct the shape of the band (2,1) absorbing from the 
first excited vibrational levels 
a higher value must be assumed, i.e. above 750 K.
Thus the excitation conditions of the TiO bands are similar to those in 2005
\citep[see][]{keckI}, indicating that the bands showing a high blueshift are
formed well above the photosphere.
A column density estimated from the TiO $\gamma$ (1,0) band is now a factor
of 2 lower than in 2005.

\subsubsection{ScO}

The shape of two components of the ScO A$^{2}\Pi$-X$^{2}\Sigma^{+}$ (0,0)
band is very similar to that observed in 2005.
The simulations show that
one cannot derive a single value of the excitation temperature from the observed contour.
The head formed by lines absorbing
from highly excited levels, i.e. 
 $^R$R$_{1G}$ at 6064.31\,\AA, 
needs an excitation temperature above 750\,K, and
is formed at a velocity of 66~\kms\ (45~\kms\ in October 2005).
The dominant round absorption contours at  6042 and 6082~\AA, formed
by the Q and P branches, indicate velocities as low as --10~\kms\ and an excitation
temperature below 350~K. The features of the A -- X (0,0) band,
formed by the $^{R}$R$_{2G}$+$^R$Q$_{2G}$ and $^Q$Q$_{1G}$+$^Q$R$_{1G}$ heads, 
have velocities of about --55~\kms. 
This velocity component was also observed in the 2005 spectrum, but now it
indicates a lower column density.
An additional velocity component at --30~\kms\ can
be recognized in the discussed heads in the present spectrum. 

The ScO A -- X (0,1) band, observed in emission at 6410 and 6460\,\AA\ in
2005, is not seen in the present spectrum, neither in emission nor in absorption.

\subsubsection{YO}

The YO subband of A$^{2}\Pi$-X$^{2}\Sigma^{+}$ (0,0) at 5975~\AA\ 
is well reproduced when assuming an excitation temperature in the range of 
350 -- 500~K. The velocity of the absorbing gas is then about
16~\kms. Similarly, from the fit to the second subband at
6135\,\AA\ one can infer an
excitation temperature of $\sim$350 K and a velocity of 36~\kms.
Both the excitation temperature and the column density are similar to 
those obtained from the 2005 spectrum \citep{keckI}. 
However, in 2005, the YO band components were observed at a velocity of 
--8~\kms.

\subsubsection{VO}

In the observed spectral range only vibrational bands of the VO
C$^4\Sigma^{-}$ -- X$^4\Sigma^{-}$ electronic system are observed.
In two bands, i.e. (0,0) at 5735~\AA\ and (1,0) at 5468~\AA, absorbing 
from the ground level,
three velocity components can be inferred from their shapes. 
The most blueshifted narrow feature is at --74~\kms.
A less clear and dispersed in velocity component is observed in a range of 
--20 to --40~\kms. Another component dominating the redward part of the band has
a velocity of 20~\kms.
A less intense VO C-X (2,0) head is observed at a velocity of about 6~\kms, 
but here also an extension of absorption towards negative velocities is
evident.

In 2005, a velocity component of --77~\kms was observed 
in the (1,0) and (0,0) bands \citep{keckI}. Within the limits of
uncertainity we can identify it with the --74~\kms\ component in the present
spectrum. 
From our simulations of the (0,0) band in 2009, we can infer
an excitation temperature of $\sim$300~K, i.e. a similar value as that
obtained in 2005. However, the column density in 2009 decreased by at least a
factor of 2 compared to the result from the 2005 spectrum.

\subsubsection{AlO  \label{alo_sect}}

\begin{figure}
  \resizebox{\hsize}{!}{\includegraphics{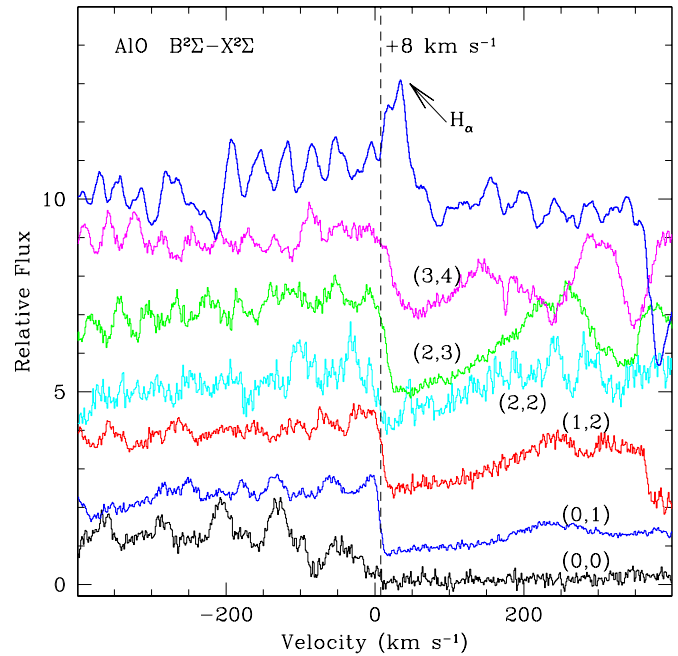}}
  \caption{
Detailed view of the AlO B$^{2}\Sigma$ -- X$^2\Sigma$ band-heads
in the velocity scale (heliocentric frame). The characteristic velocity of 8~\kms\
is marked with the vertical dashed line. 
The H$\alpha$ emission is also shown at the top of the figure.
}
  \label{FigAlO}
\end{figure}

Contrary to TiO, the observed AlO bands have
generally very sharp heads if they are absorbing from excited vibrational
levels. 
A sample of the AlO band heads is shown in Fig.~\ref{FigAlO}. Our analysis
of them can be summarized as follows.

\begin{itemize}
\item{
In the spectral range of the present spectrum there is only one band of 
AlO B$^{2}\Sigma$-X$^{2}\Sigma$ originating from the ground vibrational level, 
(0,0), at 4842\,\AA.
The head of this band is blended with TiO band rotational lines.
Apparently, the band-head extends in velocity from the minimum of absorption 
at 16 down to at least --100~\kms.
}
\item{
All AlO B-X bands originating from excited vibrational levels
have sharp band-heads at $\sim$8~\kms, with a tendency to shift up to
$\sim$25~\kms\ with the increasing vibrational number of the lower level.
}

\item{
A regular pattern of the rotational transitions formed
by blends of two main P-branches can be easily recognized in the band contour.
A good example is the AlO (0,1) band at 5079\,\AA. The radial velocity obtained 
from fitting the synthetized spectrum amounts to 52~\kms.
With the increasing rotational number J a new component with a velocity of 
$\sim$83~\kms\ appears, overlapping the blueshifted lines.
This pattern repeats in the absorptions from the v"=2 level.
Above v"=2 the rotational structure is consistent with
a velocity of 88~\kms. 
}

\item{
Redward from the band-head, the rotational pattern seems to be smoothed out,
either by
a velocity gradient, or by a high density of transitions of the two R-branches,
or by both mechanisms.
}
\end{itemize}

The rotational contours of the AlO bands are relatively well reproduced if 
the excitation temperature is of $\sim$1500\,K.
A similar value was obtained from the AlO bands in the 2005 spectrum
\citep{keckI}.
The most significant difference between 2005 and 2009 is
the position of the heads of the excited vibrational bands. The present value
of 8~\kms\ is to be compared with --82~\kms\ in 2005.

\subsubsection{Summary}

The main conclusions that we drawn from the above description of the
molecular bands are:

\begin{itemize}
\item{The excitation temperature of the molecular bands of all observed
molecules in 2009 is consistent within the uncertainties
with the values derived from the 2005 spectrum.
}
\item{Differences in the shape of the molecular bands observed in the two
epochs result mainly from the lower column density in 2009, 
typically by a factor of 2.
}
\item{The kinematic structure as inferred from the heads of the molecular 
bands is generally similar in both epochs.
}
\item{Exceptional are the highly excited lines, which are formed at 88\,km\,s$^{-1}$ 
in 2009, compared to 57\,km\,s$^{-1}$ observed in 2005.
}
\item{The change in the velocity is most spectacular in the excited AlO bands.
In 2009, these heads are at 8\,km\,s$^{-1}$, whereas in 2005 the head
velocities were --89\,km\,s$^{-1}$.
}
\end{itemize}

\section{Conclusions and discussion}
Globally, V838 Mon did not significantly evolve between 2005 and 2009. It is now slightly more 
luminous and hotter than it was in 2005
(Sect.~\ref{sed}). It continues to lose matter although at a lower rate. This is
indicated by the disappearance of clear emission components in the
atomic lines showing P-Cyg profiles in 2005 (Sect.~\ref{profiles}).

Several atomic absorption lines display a narrow and faint emission
component. Our analysis shows that this is a real emission, most probably
a remnant of the P-Cyg emission component observed in 2005, and that its
position measures the radial velocity of the object
(Sect.~\ref{em_comp}). The peak of the feature is well defined and in all
lines, where it is clearly seen, it gives $V_{\rm h} = 71$~\kms\
(equivalent to $V_{\rm LSR} = 54$~\kms),
the same value as that of the SiO masers \citep{deguchi,claussen}.
A very similar radial velocity was also measured for a molecular region seen in a
near vicinity of V838~Mon in the CO rotational lines \citep{kdt10}.
The region is likely to be associated to the dusty matter responsible for the light
echo of the 2002 eruption.

An absorption
component at radial velocities positive in respect to the systematic
velocity of the object (Sect.~\ref{sect_profiles} and Fig.~\ref{profiles})
is one of the most common features in the atomic lines. 
This component was also observed in 2005 but was
less evident and seen in fewer lines. A likely reason could have been the strong P-Cyg
emission component in 2005, which could have filled up the redward absorption in
certain lines. But it can also be argued that the redward absorption component
became intrinsically stronger in 2009. This is supported by the
appearance of a component at 88~\kms\ in the highly excited bands of TiO and
other molecules in 2009 (Sect.~\ref{bands}). 

Similarly as in \citet{keckII}, we interpret the redward absorption
component as evidence of infall motions in the atmosphere of V838~Mon. In
the present spectrum the maximum of absorption is typically observed at
$V_{\rm h} \simeq 87$~\kms, i.e. at an infall velocity of $\sim$16~\kms. The
long wavelength wing of the feature extends to $V_{\rm h} \simeq 120$~\kms,
indicating a maximum infall velocity of $\sim$50~\kms. This is consistent
with the free-fall velocity for an 8~$M_\odot$ star \citep{tss05} 
at a photospheric radius of $\sim400~R_\odot$ (Sect.~\ref{sed}), which is
$\sim$90~\kms. The infall motions apparently
became more intense in 2009. Note that the observed infalls cannot be
interpreted as a shrinkage of the V838~Mon remnant. With a velocity of
16~\kms\ it would result in a collapse of the effective radius by a factor
of 2 in the course of three months.

The observed kinematics of the AlO bands from the excited levels
(sharp heads at $V_{\rm h} \simeq 8$~\kms\ -- see Sect.~\ref{alo_sect})
is not seen in 
the other molecular bands nor in the atomic line profiles.
This suggests that the distribution of aluminium oxide in the
photosphere and/or the wind is different from the distributions of the other
molecules and atoms.
We calculated distributions of Al-bearing molecules in a model stellar
atmosphere taken from the MARCS database \citep{MARCS}
of an effective temperature of 3200~K and log~$g=0.0$. In the computation of
the equilibrium composition, equilibrium constants from \citet{tsuji} were
used, except AlO and AlO$_2$, for which the data were taken from a parametrization of
\citet{sharp} of the JANAF data \citep{janaf}. The results reveal that
 in the outer layers of the atmosphere, Al is mainly in the atomic phase. AlO is 
more than two orders of magnitude less abundant than the atomic Al, and
about two orders of magnitude more abundant than  AlOH and AlH.\footnote{The differences 
with previous computations \citep[e.g.][]{tenziu}
that predicted a much lower concentration of AlO are caused by
the newer equilibrium constants used in our modelling.}
A possible explanation of the observed AlO structures is that AlO is
additionally produced and exited in processes of non-equlibrium chemistry 
in a postshock region. This would explain both
the presence of the sharp heads and the absorptions from the excited vibrational levels, 
if the postshock temperature and density were sufficiently high.
The position of H$\alpha$ emission, as seen in
Fig.~\ref{FigAlO}, with its blue edge at
$\sim$8 km\,s$^{-1}$ could also have been explained in this hypothesis.

Absorption features showing the highest outflow velocities, i.e. $V_{\rm h}$ 
between $-130$ and $-50$~\kms, are observed in the
molecular bands arising from the ground vibrational level 
(Sect.~\ref{bands} and Table~\ref{TableMolec}) and in some atomic lines, mostly in
the resonance ones (Fig.~\ref{profiles}).
With the
systematic velocity of the object ($V_{\rm h} = 71$~\kms), these features
imply outflow velocities of 120--200~\kms.
The structures of these high outflow-velocity molecular bands usually
give very low excitation temperatures, i.e. of 300--500~K
(Sect.~\ref{bands}). 

In 2005, a relatively narrow absorption component (NAC)
at $V_{\rm h} \simeq -82$~\kms\ was observed in \ion{Rb}{I}\,\lam\,7800 
and \ion{Mg}{I}\,\lam\,4571 \citep[see Fig.~6 in][]{keckI}. 
These two lines are, unfortunately, outside the spectral range of the 2009 spectrum.
However, a similar
component at practically the same velocity is now observed in \CaI\,\lam\,6572 
and \LiI\,\lam\,6708 (see panel a in Fig.~\ref{profiles} and
Fig.~\ref{monthly}). 

In \citet{keckII}, all features in the atomic lines
were intepreted in terms of a standard spherically symmetric wind model. 
Within this model, the NAC
would have originated from a local density maximum in the wind, 
produced by a short-lived high mass-loss event.
A jet-like structure in the ongoing mass outflow was considered as another
possible explanation of the NAC.
A third possibility  was that the NAC had
nothing to do with the current mass outflow from V838~Mon, but instead it
was formed in a distant matter ejected by V838~Mon during the 2002 eruption.
This last explanation was rejected mainly on an argument raised 
from a discussion of the \TiI\ lines. That argument was incorrect, however,
since the excitation energy of the upper level of the \TiI\ lines was
mistaken as that of the lower level in \citet{keckII}. Below we show that
the 2002 ejecta is the only reasonable explanation not only of the NAC, but
probably of all the absorption features observed at high outflow velocities.

A short-lived mass-loss event in the ongoing mass outflow can immediately be
rejected in view of Sect.~\ref{var} and Fig.~\ref{monthly}. In the standard
wind model of \citet{keckII}, the shell producing the NAC was at
$\sim$2~radii of the central star. Therefore, the feature should have
evolved on a time scale of $\sim$20~days.
The features at 
$V_{\rm h} \la -50$~\kms\ in the \CaI\ and \LiI\ lines were very stable over 
a time span of almost three months. A jet-like outflow could in principle
produce a stable feature at the projected velocity of the jet. However, it
is not clear how it could produce such a complex and stable structure as that 
observed in the \LiI\,\lam\,6708 (condensations in the jet would flow out on
a time scale of days). Besides, as noted above, the molecular bands
displaying high outflow velocities show very low excitation temperatures, so
they cannot be formed close to the photosphere of V838~Mon.

During the time span between our individual observations in 2009, 
matter flowing with a velocity of
$\sim$150~\kms\ covered a distance of $\sim$1500~$R_\odot$, i.e.
$\sim$4~radii of the V838~Mon remnant (see Sect.~\ref{sed}). Therefore, the
only reasonable way of explaining the observed stability of the high outflow
velocity features in the \CaI\,\lam\,6572 and \LiI\,\lam\,6708, as well as of
the narrow absorptions at $V_{\rm h} = -92$~\kms\ in the \BaI, \BaII, and \TiI\ lines 
(see panels b and c in Fig.~\ref{profiles}), is to conclude that the above
estimated distance of $\sim$1500~$R_\odot$ is negligible compared to the distance of
matter in which the features are formed. On the other hand, comparing the
line profiles observed in 2009 with those observed in 2005, we see that the
discussed features evolve on a time scale of years. The conclusion is quite
clear: the high-velocity features are formed in the matter ejected by
V838~Mon in 2002. During the 2002
eruption the observed outflow velocities were between
$\sim 100$ and $\sim 600$~\kms\ \citep{muna02,crause03,kipp04}.
The most intense mass loss probably took place at the end of the
eruption, i.e. in March~2002 \citep{tyl05}, with outflow velocities
ranging from $\sim$100 to $\sim$300~\kms\ \citep{crause03,tyl05}. 
In the 2005 and 2009 spectra, we thus observe relatively slow but dense
parts of the 2002 ejecta. Faster and presumably less dense regions quickly
became too thin to be observable.

The above hypothesis follows \citet{lynch04,lynch07}, who concluded from
their earlier observations that strong molecular bands observed in the
near-IR were formed in the expanding 2002 ejecta. 
It explains several observational facts from our observations that are not
easily understood otherwise, i.e. abnormaly strong molecular bands in the spectrum of 
V838~Mon when compared to typical giants or supergiants of a similar
spectral type, very low excitation temperatures of the molecular bands
formed at high outflow velocities, and decrease in
the column densities observed in the high velocity molecular bands 
between 2005 and 2009 (Sect.~\ref{bands}). It is now clear also why the
maximum observed outflow velocity of 215~\kms\ was the same in 2005 and 2009
and removes problems with explaining that high value in terms of the winds
from cool giants \citep{keckII}.

From the low-excitation TiO and VO bands observed in the October~2005, 
\citet{keckI} derived a column density of H atoms of $\sim$24~dex~cm$^{-2}$,
i.e. $\sim$2~g\,cm$^{-2}$. If the matter had been lost in March~2002 with a
velcity of 150~\kms, it would have been at a distance of $\sim 2.5 \times 10^4~R_\odot$ 
in October~2005. Assuming spherical symmetry, one obtains a mass of the
expanding shell of $\sim 0.04~M_\odot$. This result agrees well with
an estimate of the expanding mass made in \citet{lynch04} and is within the
limits of the mass lost by V838~Mon in the 2002 eruption obtained in
\citet{tyl05}.

The disappearance of the B3~V companion from the spectrum of V838~Mon is the
most intriguing event in the recent years' history of the object. It started with an
eclipse-like event in November--December 2006 \citep{muna07}. Then the
companion reappeared in the spectrum for a few months and next started a
slower but deeper and more complex decline. Note that
after the 2002 eruption the $U$ brightness of V838~Mon was dominated by the
B3~V
companion. Our spectroscopy shows that in 2009 the B3~V companion was at
least 30 times fainter in the optical than in 2005
(Sect.~\ref{spect_features}). Following the interpretation of
\citet{keckII}, we can conclude that the B3~V companion is now engulfed in a
dense dusty cloud of matter lost by V838~Mon during the 2002 eruption. The
luminosity of the companion is thus expected to be now reemitted in the
infrared. Therefore, one should look for traces of the companion in IR
observations. This can appear to be a hard task, however, since the luminosity of the
B3~V companion was only of a few per cent of that of the V838~Mon remnant.

Given that the B3~V companion has already disappeared for a few
years, one can speculate that in 2007 the companion entered the
V838~Mon remnant and is now completely engulfed in the remnant
itself. This idea would imply a second merger phase in the history of V838~Mon,
in which the B3~V companion would finally merge with the core of V838~Mon.
This would result in a much more energetic and dramatic event than the 2002 eruption.
We do not follow this speculation any further, as it does not seem very
likely in view of the observational facts. 
Since its discovery in October~2002 until the eclipse-like event in
2006, the spectrum of the B3~V
companion showed no interactions of the star with the 2002 ejecta. 
In particular, the $U$ brightness, dominated by the B3~V companion in that
epoch, remained virtually constant.$^1$
This shows
that the companion is at a significant distance from V838~Mon, so that the
matter ejected in 2002 was not able to interact with the companion until it
reached the star at some point in 2006.

Future observations will distinguish between the two scenarios. It is very
likely that V838~Mon has not yet finished to suprise us.
Undoubtedly, the object deserves further observational monitoring.

\acknowledgements{R.T., T.K., M.S., and T.T. acknowledge support from 
the Polish Ministry of Science and Higher Education under grants
N\,N203\,403939 and N203\,018\,32/2338. R.K. acknowledges support from 
Cento de Astrof\'isica de Valpara\'iso and Proyecto DIPUV 23/2009. 
We thank K. He\l{}miniak, P. Pietrukowicz, M. Ratajczak, 
and the TCS staff  for obtaining the photometric observations reported in this paper.
We acknowledge the use of data from the UVES Paranal Observatory Project
(ESO DDT Program ID 266.D-5655).
}

\bibliographystyle{aa}

\begin{appendix}
\section{Molecular bands in the spectrum of V838~Mon observed in 2009}
\label{table_molec}

A detailed identification of the molecular features in the spectrum of
V838~Mon was made in \citep[][see their Table~3]{keckI}. The list of the
molecular bands observed in 2009 remains practically the same as in 2005.
Table~\ref{TableMolec} presents details of those molecular bands observed in
the spectrum of V838~Mon in 2009 for which we were able to estimate the
radial velocity. First column displays the observed
wavelengths of the bands, while their laboratory wavelengths are given in
the second column.
Identification details of the bands are presented in the third column.
Heliocentric radial velocities of the bands are given in the fourth
column. Three different methods were used to derive the velocities, as
indicated in the last column of the table. The meanings of the symbols are as
follows:  M5 III -- comparison of the V838~Mon spectrum to the UVES spectrum of 
an M5 III star HD118767 \citep{paranal};
LAB -- direct comparison of the band head position with the laboratory
measurements \citep[see][for laboratory data sources]{keckI};
SYNTH - spectral synthesis of the bands as described in \citet{keckI}.

\longtab{1}{

\begin{longtable}{lllcl}
\caption{\label{TableMolec} Heliocentric radial velocities of the molecular bands in the
 2009 spectrum of V838~Mon} \\
\hline\hline
$\lambda_{\rm{obs}}$ & $\lambda_{\rm{lab}}$ & Identification & Velocity & Method \\
(\AA) & (\AA) & & (km s$^{-1}$) & \\
\hline
\endfirsthead
\caption{continued.}\\
\hline\hline
\renewcommand{\footnoterule}{}
$\lambda_{\rm{obs}}$ & $\lambda_{\rm{lab}}$ & Identification & Velocity & Method \\
(\AA) & (\AA) & & (km s$^{-1}$) & \\
\hline
\endhead
\hline
\endfoot
4788-4804 &           & TiO $\alpha$ (2,0)  high J tail &  +88$\pm$5   & M5 III \\
4805.2    & 4804.333  & TiO $\alpha$ (3,1)  R$_2$       &  +56$\pm$25  & LAB    \\
4831-4841 &           & TiO $\alpha$ (3,1)  high J tail &  +88$\pm$5   & M5 III \\
4842.47   & 4842.27   & AlO B$^{2}\Sigma$-X$^{2}\Sigma$ (0,0) R$_2$ & +12$\pm$10 & LAB \\
4889.26   & 4889.0    & AlO B$^{2}\Sigma$-X$^{2}\Sigma$ (2,2) R$_2$ & +16$\pm$10 & LAB \\
4953.0$\pm$0.2 & 4954.592 &  TiO $\alpha$ (1,0)         &  -90$\pm$18  & LAB    \\
4993-4998 &           & TiO $\alpha$ (1,0) high J tail  &  +88$\pm$10  & M5 III \\
5042-5078 &           & TiO $\alpha$ (3,2)              &  +88$\pm$10  & M5 III \\
5079.5    & 5079.36   & AlO B$^{2}\Sigma$-X$^{2}\Sigma$ (0,1)                 &   +8$\pm$5   & LAB    \\
5087-5096 &           & AlO B$^{2}\Sigma$-X$^{2}\Sigma$ (0,1)  P-branch       &  +52$\pm$5   & SYNTH  \\
5102.3    & 5102.13   & AlO B$^{2}\Sigma$-X$^{2}\Sigma$ (1,2)  R-head           &  +10$\pm$5   & LAB    \\
5113-5118 &           & AlO B$^{2}\Sigma$-X$^{2}\Sigma$ (1,2)  P-branch         &  +52$\pm$5   & SYNTH  \\
5123.6    & 5123.33   & AlO B$^{2}\Sigma$-X$^{2}\Sigma$ (2,3)  R-head           &  +15$\pm$6   & LAB    \\
5143.3    & 5142.89   & AlO B$^{2}\Sigma$-X$^{2}\Sigma$ (3,4)  R-head           &  +23$\pm$10  & LAB    \\
5130-5137 &           & AlO B$^{2}\Sigma$-X$^{2}\Sigma$ (3,4)  high-J           &  +88$\pm$10  & SYNTH  \\
5164.4    & 5166.450  & TiO $\alpha$ (0,0) R$_{2}$      & -119$\pm$10  & LAB    \\
5197-5203 &           & TiO $\alpha$ (0,0) high-J       &  +88$\pm$10  & LAB    \\
5213-5225 &           & TiO $\alpha$ (0,0) high-J       &  +88$\pm$10  & LAB    \\
5226.3    & 5228.2    & VO  C-X (2,0)  R-head             & -109$\pm$10  & LAB    \\
5228.0    & 5228.2    & VO  C-X (2,0)  R-head             &  +11$\pm$10  & LAB    \\
          & 5240.463  & TiO f$^{1}\Delta$-a$^{1}$ (0,0)   &  +88$\pm$10  & M5 III \\
          & 5259.418  & TiO $\alpha$ (2,2) R$_{2}$      &  +88$\pm$10  & M5 III \\
          & 5275.8    & VO C-X (3,1)                      &  +88$\pm$10  & M5 III \\
          & 5307.116  & TiO $\alpha$ (3,3)              &  +88$\pm$10  & M5 III \\
5377.21   & 5376.813  & AlO B$^{2}\Sigma$-X$^{2}\Sigma$ (2,4) & +22$\pm$10 & LAB  \\
5467.7    & 5469.3    & VO C-X (1,0)                    &  -88$\pm$10  & LAB    \\
5468.3    & 5469.3    & VO C-X (1,0)                    &  -52$\pm$10  & LAB    \\
5488-5498 &           & TiO $\alpha$ (0,1) high-J       &  +88$\pm$10  & M5 III \\
          & 5496.423  & TiO $\alpha$ (1,2) R$_2$      &  +88$\pm$10  & M5 III \\
5512-5525 &           & TiO $\alpha$ (1,2) high-J     &  +88$\pm$10  & M5 III \\
          &           & TiO $\beta$ (0,0)               &  +88$\pm$10  & M5 III \\
          &           & TiO $\beta$ (1,1)               &  +88$\pm$10  & M5 III \\
          &           & TiO $\beta$ (2,2)               &  +88$\pm$10  & M5 III \\
5650-5660 &           & TiO $\gamma$ (3,1) high-J tail  &  +88$\pm$10  & M5 III \\
          & ?5712.6   & TiO $\gamma$' (4,2)             &  +88$\pm$10  & M5 III \\ 
          & ?5717.0   & TiO $\gamma$' (4,2)             &  +88$\pm$10  & M5 III \\  
5735.22   & 5736.703  & VO C-X (0,0)                    &  -78$\pm$10  & LAB    \\
5735.98   & 5736.703  & VO C-X (0,0)                    &  -38$\pm$10  & LAB    \\
          & 5758.741  & TiO $\alpha$ (0,2) R$_2$      &  +88$\pm$10  & M5 III \\
5845.73  & 5847.593  & TiO $\gamma$' (1,0) R$_1$       &  -97$\pm$10  & LAB    \\
5950      &           & TiO $\gamma$' (2,1) R$_{3}$     &  +88$\pm$10  & M5 III \\
5999.64   & 6001.151  & TiO$\gamma$ (3,0) R$_{2}$       &  -75$\pm$10  & LAB    \\
          & 6064.31   & ScO (0,0) R$_{1}$               &  +88$\pm$10  & M5 III \\
6077.65   & 6079.30   & ScO (0,0) Q$_{1}$               &  -81$\pm$20  & LAB    \\
6147.15   & 6148.68   & TiO $\gamma$' (0,0) $^S$R$_{21}$ &  -75$\pm$10 & LAB    \\
6269.3    & 6268.86   & TiO $\gamma$' (1,1)             &  +18$\pm$10  & M5 III \\
6292.2    & 6294.80   & TiO $\gamma$ (2,0) R$_3$        &  -123$\pm$20$^a$ & LAB    \\
6294.2    & 6294.80   & TiO $\gamma$ (2,0) R$_3$        &  -29$\pm$10$^a$  & LAB    \\
6296.0    & 6294.80   & TiO $\gamma$ (2,0) R$_3$        &  +57$\pm$5$^a$  & LAB    \\
6318.9    & 6321.21   & TiO $\gamma$ (2,0) R$_2$        & -110$\pm$10$^a$  & LAB    \\
6320.7    & 6321.21   & TiO $\gamma$ (2,0) R$_2$        &  -24$\pm$10$^a$  & LAB    \\
6349.7    & 6351.29   & TiO $\gamma$ (2,0) R$_1$        &  -75$\pm$20$^a$  & LAB    \\
6398-6412 &           & TiO $\gamma$ (3,1) high-J tail &  +88$\pm$10  & M5 III \\
          & 6447.90   & TiO $\gamma$ (4,2)              &  +88$\pm$10  & M5 III \\
          & 6478.55   & TiO $\gamma$ (4,2) R$_1$        &  +88$\pm$10  & M5 III \\
          & 6512.409  & TiO $\gamma$ (5,3) R$_2$        &  +88$\pm$10  & M5 III \\
          & 6543.950  & TiO $\gamma$ (5,3) R$_1$        &  +88$\pm$10  & M5 III \\
          & 6549.600  & TiO $\gamma$ (6,4) R$_3$        &  +88$\pm$10  & M5 III \\
          & 6577.716  & TiO $\gamma$ (6,4) R$_2$        &  +88$\pm$10  & M5 III \\
6649.34   & 6651.252  & TiO $\gamma$ (1,0) R$_{3}$      &  -86$\pm$20  & LAB    \\
6678.81   & 6680.796  & TiO $\gamma$ (1,0) R$_{2}$      &  -89$\pm$20  & LAB    \\
6712.59   & 6714.477  & TiO $\gamma$ (1,0) R$_{1}$      &  -84$\pm$20  & LAB    \\
          & 6717.578  & TiO $\gamma$ (2,1) R$_{3}$      &  +68$\pm$5   & M5 III \\
6733-6740 &           & TiO $\gamma$ (1,0) high-J tail  &  +68$\pm$5   & M5 III \\
          & 6747.589  & TiO $\gamma$ (2,1) R$_{2}$      &  +68$\pm$5   & M5 III \\
6765-6780 &           & TiO $\gamma$ (2,1) high-J tail  &  +88$\pm$5   & M5 III \\
6808-6815 &           & TiO $\gamma$ (2,1) high-J tail  &  +88$\pm$5   & M5 III \\
\hline

\end{longtable}
$^a$No clear bandhead, a velocity gradient in the profile is observed.

}
\end{appendix}

\end{document}